\documentstyle[epsf,wrapft]{ptptex}

\notypesetlogo  

\newcommand{\vect}[1]{\mbox{\boldmath${#1}$}}
\newcommand{\s}{\hspace{3mm}}

\title{
  A Study of the Thermodynamic Properties of a Hot, Dense Hadron
  Gas Using an Event Generator
}

\subtitle{
  {\it Hadro-Molecular-Dynamical Calculation }
}

\author{
  Nobuo {\sc Sasaki}
}

\inst{
  {\it Physics Department, Hiroshima University,\\ 
           Higashi-Hiroshima 739-8526, Japan}
}

\recdate{
  April 30, 2001
}

\abst{
  We investigate the equilibration and the equation of state of a
  hot hadron gas at finite baryon density using an event generator
  which is designed to approximately satisfy detailed balance at
  finite temperatures and finite baryon densities of the hadronic scale
  ($80$ MeV $< T < 170$ MeV and $0.157$ fm$^{-3} < n_{B} < 0.315$
  fm$^{-3}$).  Molecular-dynamical simulations were performed for
  a system of hadrons in a box with periodic boundary conditions.
  Starting from an initial state composed of only nucleons with a
  uniform phase-space distribution, the evolution takes place through
  interactions such as collisions, productions and absorptions.  The
  system approaches a stationary state of baryons, mesons and
  their resonances, which is characterized by the value of the
  exponent in the energy distribution common to the different
  particles, i.e., the temperature.  After the equilibration,
  thermodynamic quantities, such as the energy density, particle
  density, entropy and pressure are calculated.  Above
  $T \sim m_{\pi}$, the obtained equation of state exhibits a
  significant deviation from the naive mixed free gas model.  
  Large values of the entropy per baryon are also notable.
  In our system, the excitation
  of heavy baryon resonances and meson production are enhanced
  simultaneously, and the increase of the temperature becomes
  moderate, but a Hagedorn-type artificial temperature saturation
  does not occur.  The pressure exhibits a linear dependence on
  the energy density.
}

\begin{document}

  \maketitle   

  \section{Introduction}
  \label{sec:intro}
    From the viewpoint of microscopic dynamics, a hot, dense hadron gas
    has many unique properties, e.g., the coexistence of light
    relativistic particles and heavy non-relativistic particles, the
    production and absorption of particle--anti-particle pairs, and a
    large number of degrees of freedom of resonance states.  Such
    features are quite different from those of ordinary classical
    molecular dynamical systems, and properties of hadron gases in
    on-equilibrium and off-equilibrium are highly nontrivial and a
    very interesting topic for theoretical study.  In particular the
    equation of state (EOS) and transport coefficients of hot, dense
    hadron gases are quite important quantities both in high-energy
    nuclear physics and cosmology.  In the ultra-relativistic
    heavy-ion experiments at CERN and BNL, though the primary purpose
    is the search for a quark-gluon plasma (QGP), the physics of a
    hadron gas dominates the resulting system, and knowledge of the
    EOS and transport coefficients of a hadron gas is highly necessary
    for a better understanding of the experimental
    results.\cite{ref:QM99} \ In cosmology, inhomogeneous
    nucleosynthesis at an early stage of the universe is considered
    one of the possibilities for the origin of matter, and the baryon
    diffusion constant is a key quantity in this
    scenario.\cite{ref:Kaji1}

    In spite of their great importance, the EOS and transport
    coefficients of hot, dense hadron gases are still poorly known,
    because of the nonperturbative nature of the strong interaction.
    Numerical simulations based on lattice QCD represent a powerful
    tool for non-perturbative QCD, and many calculations of EOS for
    hot quark-gluon plasma phase have been carried
    out.\cite{ref:Kar1}\tocite{ref:Boyd1} \ However, quantitative
    investigation of the hadron phase is very difficult, and only few
    useful results have been obtained.  In spite of several novel
    approaches,\cite{ref:Bar1,ref:Eng1} \ inclusion of the chemical
    potential is still difficult for {\it SU}(3) lattice QCD.
    Moreover, progress in the study of transport coefficients is very
    slow, and only a calculation for the pure gluonic matter has been
    completed.\cite{ref:Saka1}

    Phenomenologically, many thermodynamic models have been proposed.
    Though these models can describe some aspects of the properties
    of the hadronic matter, whether they are realistic enough or not
    is unclear.  For example, Hagedorn proposed a bootstrap model
    many years ago, but the problem of the limiting temperature has
    been pointed out.\cite{ref:Hage1} \ A hot, dilute hadron gas is
    often regarded naively as an ideal gas of massless pions and
    massive baryons.  Since a pion is the lightest mode to be
    predominantly excited, this picture seems to provide a reasonable
    starting point.  However, residual interactions may not be
    negligible.  This simple picture can be partly improved by taking
    the ``size'' of hadrons into account through an excluded volume
    effect, as in the van der Waals
    equation.\cite{ref:Gore1}\tocite{ref:Kou1} \ However, it is still
    not clear how appropriate such an approximation is.  Although the
    excluded volume imitates the effect of the repulsion between
    hadrons, microscopic interactions of a hot, dense hadron gas are
    much more complicated, and their roles are not trivial.  Thus, we
    need to investigate the thermodynamic properties of a hadron gas
    by using a microscopic model that includes realistic interactions
    among hadrons.  In this work, we have adopted a relativistic
    collision event generator, URASiMA (ultra-relativistic AA collision
    simulator based on multiple scattering
    algorithm),\cite{ref:Kino1}\tocite{ref:Kuma1} \ and performed
    molecular-dynamical simulations for a system of a hadron gas.

    In this paper, we focus our interest on the ``hadronic scale''
    temperature ($80$ MeV $< T < 170$ MeV) and baryon number density
    ($0.157$ fm$^{-3} < n_{B} < 0.315$ fm$^{-3}$), which are expected
    to be realized in high energy nuclear collisions.  Thermodynamic
    properties and transport coefficients of hadronic matter in this
    region should play important roles in phenomenological models.
    Sets of statistical ensembles are prepared for the system of
    fixed energy density and baryon number density.  Using these
    ensembles, the equation of state is investigated in detail.  The
    statistical ensembles have already been applied to calculations
    of the diffusion constant of a hot, dense hadron
    gas.\cite{ref:Sasa2} \ Calculations of viscosity and heat
    conductivity are also now in progress.\cite{ref:Sasa3}

    In a previous work, we made a pilot study of the equation of
    state of a hot, dense hadron gas with the event generator
    URASiMA.\cite{ref:Sasa1} \ In that work, an approximate saturation
    of the temperature, which behaves like the Hagedorn limiting
    temperature, appears.  Recently, Belkacem et al. performed a
    similar calculation using a different event generator, UrQMD,
    and they also reported the saturation of the
    temperature.\cite{ref:Bass1,ref:Bel1} \ Though those works have
    provided valuable information regarding the nature of the hadron
    gas, some results are misleading, because, in those simulations,
    the detailed balance of the processes is explicitly broken.
    Neither model includes multi-body absorptions, which are reverse
    processes of multi-particle production.  For energetic hadrons
    in a closed system, we need both processes to ensure detailed
    balance.  If detailed balance is broken, the reversibility of the
    equilibrated system is no longer realized.  A luck of the
    reversal process of multi-particle production can cause one-way
    conversion of kinetic energy into particles, and artificial
    saturation of the temperature can occur.  Although it is
    interesting and important to formulate these multi-body
    absorption processes exactly in our simulation, straightforward
    implementation of them is very difficult and not practical.
    In this work, avoiding this complicated problem, we employed a
    practical approach.  We improved URASiMA to almost recover the
    detailed balance at temperatures of present interest
    \footnote{On the order of several hundred MeV.} by adopting an
    idea that is discussed in the next section.

    In \S\ref{sec:method}, we describe the method of our simulation.
    We also introduce URASiMA and explain the method to improve it so
    that it almost recovers detailed balance effectively.  In
    \S\ref{sec:results}, we discuss the equations of state of the
    hadron gas, and compare them with those of the free gas model. 
    In order to confirm that the obtained result is insensitive to the
    system size, we also investigate the finite size effect.  Section
    \ref{sec:summary} is devoted to a summary and discussion of the
    outlook.

  \section{Tool and method}
  \label{sec:method}
    \subsection{Method of simulation}
    \label{subsec:method}
\begin{wrapfigure}{r}{5cm}
  \epsfxsize=5cm
  \epsfbox{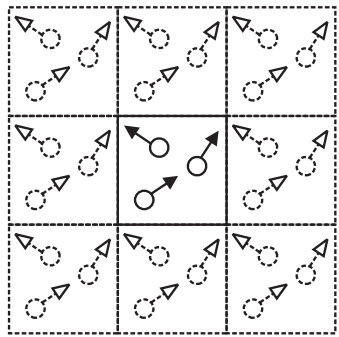}
   \caption{ 
    This picture depicts the idea of periodic boundary conditions.
    The box that is located in the center is the space of interest.
    We call this the `real box'.
  }
  \label{fig:peri}
\end{wrapfigure}
    In this work, we focus our investigation on the thermodynamic
    properties of a hadronic system.  For this purpose, we consider a
    system in a cubic box and impose periodic boundary conditions in
    configuration space; i.e., if a particle leaves the box, another
    one with the same momentum enters from the opposite side.  We
    display the boundary conditions pictorially in Fig. \ref{fig:peri},
    where the box located in the center is the system we refer to as
    the `real box' and the others are replicas.  When we survey
    would-be collision points of the particles in the real box, we
    have to survey the collisions not only between particles in the
    real box but also with the particles in replicas.

    The energy density $\varepsilon$ and the baryon number density
    $n_{B}$ in the box are fixed as input parameters, and these
    quantities are conserved throughout the simulation.  The initial
    distributions of nucleons are given by uniform random distributions
    in phase space.  The momenta of the particles are adjusted to
    satisfy the initial condition of energy, $\sum_{i=1}^{N}
    \sqrt{m^{2}_{i} + \vect{p}^{2}_{i}} = \varepsilon \cdot V$, in
    the center-of mass system of the particles:
    $\sum_{i=1}^{N} \vect{p}_{i} = \vect{0}$.  After setting the
    initial state in the above manner, time evolution as described by
    the event generator URASiMA takes place.  Though the initial
    particles are only nucleons, many mesons are produced through
    interactions.

    In order to confirm the realization of the stationarity of the
    system, we monitor particle densities and collision frequencies. We
    also check energy distributions of each particle.  According to
    these methods of confirmation, as time increases, the system seems
    to become stationary.  Whether the energy distributions approach
    the Boltzmann distribution is a fundamental question.  If the
    system is in thermal equilibrium, the slope parameters of the
    energy distributions for all particles should converge to the same
    value, i.e., the inverse of temperature.  In order to confirm this,
    we analyze the temporal evolution of the inverse slopes of various
    particles.

    Running URASiMA many times with the same input parameters and
    taking the stationary configuration in equilibrium, we can obtain
    statistical ensembles with fixed temperature and fixed baryon
    number density.  By using these ensembles, we can calculate
    thermodynamic quantities, such as the particle density, pressure,
    and so on, as functions of temperature and baryon number density.

    \subsection{Event generator URASiMA}
    \label{subsec:urasima}
    URASiMA was originally designed to simulate ultra-relativistic AA
    collision experiments. (Here we call it
    URASiMA-A.)\cite{ref:Kuma1} \ For this investigation, we improved
    URASiMA to study thermo-equilibrium systems in a box (URASiMA-B).
    In both models, collisions are realized when the distance between
    particles becomes smaller than their interaction range, which is
    defined by the relevant total cross-section.  We describe the
    fully relativistic method to search for would-be collision points
    in Appendix \ref{sec:colp}.  The fundamental processes in the
    URASiMA-A/B are 2-body elastic and quasi-elastic collisions
    between hadrons, multi-particle productions, and strong decays of
    resonances.  Hadronic experimental data are used as an input to the
    model.\cite{ref:pdg1} \ The processes which URASiMA-B includes are
    as follows:
    \begin{equation}
      NN \leftrightarrow NR,    \label{eq:prd1}
    \end{equation}
    \begin{equation}
      NN \leftrightarrow \Delta_{1232}\Delta_{1232}, \label{eq:prd2}
    \end{equation}
    \begin{equation}
      R \leftrightarrow N\pi,   \label{eq:dcyR1}
    \end{equation}
    \begin{equation}
      R \leftrightarrow R'\pi,  \label{eq:dcyR2}
    \end{equation}
    \begin{equation}
      R \leftrightarrow Nr,     \label{eq:dcyR3}
    \end{equation}
    \begin{equation}
      r \leftrightarrow \pi\pi, \label{eq:dcyr}
    \end{equation}
    \begin{equation}
      NN \rightarrow NN + \; n\; \mbox{secondaries}\; (n \geq 3).
      \label{eq:mult}
    \end{equation}
    Here $R$ and $r$ denote baryon and meson resonances, respectively.
    The list of particles in URASiMA-B is given in
    Table \ref{tab:plst}.  For the calculation of cross sections for
    quasi-2-body processes, we follow the work of Teis {\it et al}.
    \cite{ref:Teis1}\tocite{ref:Hub1} \ Cross sections of resonance
    absorptions such as $NR \rightarrow NN$ are calculated using the
    reciprocity of the S matrix. (See Appendix \ref{sec:abs}.)
    \cite{ref:Dan1}\tocite{ref:Li1} \ A detailed explanation of
    multi-particle production is given in Appendix \ref{sec:mul}.
\begin{table}[h]
  \caption{
    Baryons, mesons and their resonances included in URASiMA.
  }
  \label{tab:plst}
  \begin{tabular}{c|cccccccc} \hline \hline
    N        & $N_{938} $ & $N_{1440}$ & $N_{1520}$ & $N_{1535}$ &
               $N_{1650}$ & $N_{1675}$ & $N_{1680}$ & $N_{1720}$ \\
               \hline
    $\Delta$ & $\Delta_{1232}$ & $\Delta_{1600}$ &
               $\Delta_{1620}$ & $\Delta_{1700}$ &
               $\Delta_{1905}$ & $\Delta_{1910}$ &
               $\Delta_{1950}$ & \\ \hline
    meson & $\pi$ & $\sigma_{800}$ & $\rho_{770}$ &&&&& \\ \hline
  \end{tabular}
\end{table}

    In the case of high-energy hadron/nuclear collisions, absorption
    processes are not so important.  URASiMA-A includes direct
    multi-particle production ($n$ $\geq$ 1), and 1-$\pi$
    production/absorption via $\Delta_{1232}$. In spite of the limited
    number of processes and particle species, URASiMA-A reproduces
    the global features of the experimental data quite
    well.\cite{ref:Kuma1}

    However, in studies of thermo-equilibrium systems, whether
    detailed balance is maintained or not is an important property of
    the generator. Though the contributions of the multi-particle
    productions dominate the system at early stages of the
    non-equilibrated system, the reverse process plays an important
    role in the later, equilibration stage.  The absence of reverse
    processes leads to one-way conversion of the energy to particles
    and an artificial temperature saturation in the equilibrated
    system.\cite{ref:Sasa1,ref:Bel1} \ However, the exact treatment
    of multi-particle absorption processes is very difficult.
    In order to treat them effectively with URASiMA-B, direct 1-$\pi$
    and 2-$\pi$ productions are completely replaced by successive
    processes of quasi-elastic collisions and decays of resonances,
    and 1-$\pi$ and 2-$\pi$ absorptions through resonances are
    naturally included in the model.  For this purpose, URASiMA-B
    contains $\Delta$ and $N^{*}$ particles whose masses are up to
    2 GeV.  We successfully reproduced 1-$\pi$, 2-$\pi$ and inelastic
    cross sections of $NN$ collisions up to $\sqrt{s} = $ 3 GeV
    without direct productions.  This is the main difference between
    URASiMA-A and URASiMA-B.  For $\sqrt{s} > 3$ GeV, in order to
    obtain an appropriate total cross section, we need to take direct
    production processes into account.
    In our simulation, only at this point is detailed balance broken.
    Nevertheless, if the temperature is much smaller than 3 GeV, the
    influence of the breaking of detailed balance is negligibly small.
    For example, if the temperature of the system is 100 MeV, the
    occurrence of such a process is suppressed by a factor of
    exp($-30$), and thus the time scale to detect the violation of
    detailed balance is very much longer than the hadronic scale.

    In order to demonstrate the descriptive ability of URASiMA-B, we
    compare its results with experimental data of the high energy
    nuclear collisions at BNL/AGS in Appendix \ref{sec:exp}.

  \section{Results}
  \label{sec:results}
    \subsection{Parameters}
    \label{subsec:inp}
\begin{table}[h]
  \caption{
    Input parameters. Here $V$ is the volume of the box, 
    $n_{B}$ is the baryon number density, and $\varepsilon$ is the
    total energy density.
  }
  \label{tab:input}
\begin{tabular}{l|l|l} \hline \hline
  $V$ [fm$^{3}$] &
  $n_{B}$ [fm$^{-3}$] &
  \hspace{4.0cm} $\varepsilon$ [GeV/fm$^{3}$] \\ \hline
  $\;64.\;$ &
  $\begin{array}{l} .156 \\ .234 \\ .313 \end{array}$ &
  $\begin{array}{l@{\s}l@{\s}l@{\s}l@{\s}l@{\s}l@{\s}l@{\s}l@{\s}l@{\s}l@{\s}l}
    .250 & .300 & .313 & .370 & .463 & .556 & .625 & .648 &
    .741 & .833 & .938 \\
         & .300 &      & .370 & .463 & .556 &      & .648 &
    .741 & .833 & \\
         &      &      & .370 & .463 & .556 &      & .648 &
    .741 & .833 & \\
  \end{array}$ \\ \hline
  $\;128.\;$ &
  $\begin{array}{l} .156 \\ .234 \\ .313 \end{array}$ &
  $\begin{array}{l@{\s}l@{\s}l@{\s}l@{\s}l@{\s}l@{\s}l@{\s}l@{\s}l@{\s}l@{\s}l}
   .250 & .300 & \hspace{5.8mm} & .370 & .463 & .556 & \hspace{5.8mm} & .648 &
   .741 & .833 & \\
        & .300 &      & .370 & .463 & .556 &      & .648 &
   .741 & .833 & \\
        &      &      & .370 & .463 & .556 &      & .648 &
   .741 & .833 & \\
  \end{array}$ \\ \hline
  $\;216.\;$ &
  $\begin{array}{l} .157 \\ .231 \\ .315 \end{array}$ &
  $\begin{array}{l@{\s}l@{\s}l@{\s}l@{\s}l@{\s}l@{\s}l@{\s}l@{\s}l@{\s}l@{\s}l}
    \hspace{17.5mm} & &  & .370 & .463 & .556 & \hspace{5.8mm} & 
    .648 & .741 & .833 & \\
    & &  & .370 & .463 & .556 &  & .648 & .741 & .833 & \\
    & &  & .370 & .463 & .556 &  & .648 & .741 & .833 & \\
  \end{array}$ \\ \hline
  $\;1000.\;$ &
  $\begin{array}{l} .157 \\ .231 \\ .315 \end{array}$ &
  $\begin{array}{l@{\s}l@{\s}l@{\s}l@{\s}l@{\s}l@{\s}l@{\s}l@{\s}l@{\s}l@{\s}l}
    .250 & .300 & \hspace{5.8mm} & .370 & .463 & .556 &
    \hspace{5.8mm} & .648 & .741 & .833 & \\
         & .300 &  & .370 & .463 & .556 & & .648 & .741 & .833 & \\
         &      &  & .370 & .463 & .556 & & .648 & .741 & .833 & \\
  \end{array}$ \\ \hline
\end{tabular}
\end{table}

    Focusing our interest on the region of temperature and baryon
    number density which is expected to be produced in high energy
    nuclear collisions, we used the input parameters given in Table
    \ref{tab:input}.  Here $n_{B} = 0.156$ fm$^{-3}$ is taken as the
    baryon number density of normal nuclear matter.  Thus, the
    baryon number density in our simulation corresponds to 1 -- 2
    times larger than that of normal nuclear matter.  The total
    isospin of the system is set to zero, i.e., the number of protons
    is equal to the number of neutrons at the start.

    In this paper we generated a statistical ensemble through two
    different methods, the phase average ($200$ event) and the
    long-time average ($1$ event).  We compared the thermodynamic
    quantities obtained for the two ensembles in the case of
    $V=64.0$ fm$^{3}$, and we confirmed that ergodicity holds to
    sufficient precision.  In this paper, thermodynamic quantities
    are calculated from long-time averages, and the phase average is
    used when we study the time evolution of the system.

  \subsection{Chemical equilibration}
  \label{subsec:equil}
\begin{wrapfigure}{r}{7cm}
  \epsfxsize=7cm
  \epsfbox{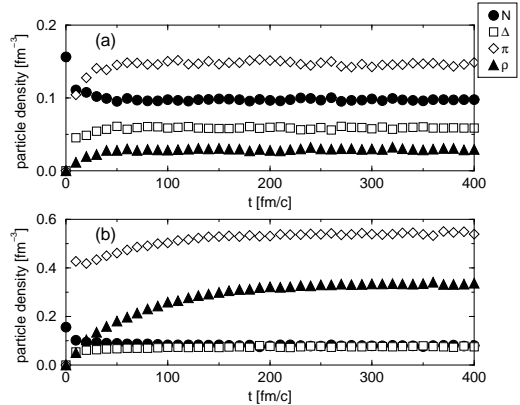}
  \caption{
    The time evolution of particle densities for each particle with
    $V = 64.0$ fm$^{3}$.  Two results are shown, (a) that for
    $n_{B}=0.156$ fm$^{-3}$, $\varepsilon=0.313$ GeV/fm$^{3}$,
    and (b) that for
    $n_{B}=0.156$ fm$^{-3}$, $\varepsilon=0.938$ GeV/fm$^{3}$.
  }
  \label{fig:pdens}
\end{wrapfigure}
\begin{figure}
  \epsfxsize=10cm
  \epsfbox{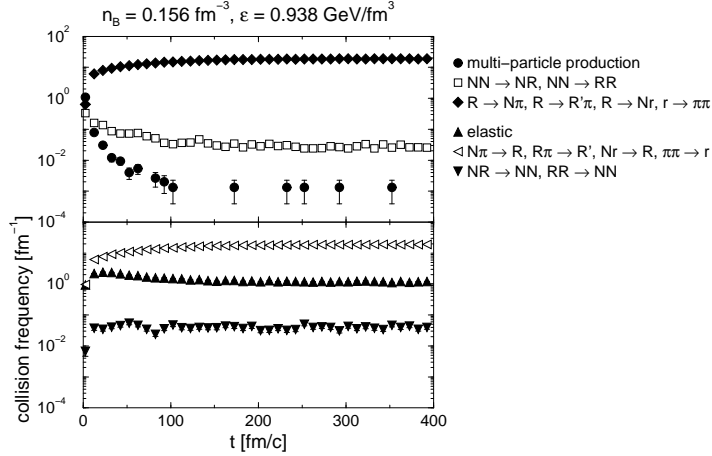}
  \caption{
    The time evolution of collision frequencies for various types of
    collisions: 2-body collisions and decays presented in
    Eqs.~(\ref{eq:prd1})--(\ref{eq:dcyr}), multi-particle productions,
    and elastic collisions. Here $R$ denotes baryonic resonances, and
    $r$ denotes meson resonances. The calculation was done with
    $V=64.0$ fm$^{3}$, $n_{B}=0.156$ fm$^{-3}$ and
    $\varepsilon=0.938$ GeV/fm$^{3}$.
  }
  \label{fig:col}
\end{figure}
\begin{figure}
  \epsfxsize=8cm
  \epsfbox{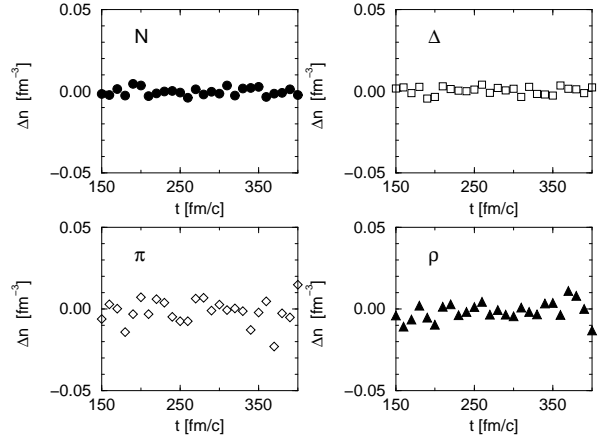}
  \caption{
    The difference between particle densities for the simulations
    with and without multi-particle production. The calculation
    without multi-particle production was realized by switching off
    the channel by hand at $t = 150$ fm/c.  The fact that the values
    in these figures are almost vanishing indicates that the
    contributions of the multi-particle productions are very small.
    The calculation was done with  $V=64.0$ fm$^{3}$,
    $n_{B}=0.156$ fm$^{-3}$ and $\varepsilon=0.938$ GeV/fm$^{3}$.
  }
  \label{fig:difnum}
\end{figure}
\begin{figure}
  \epsfxsize=8cm
  \epsfbox{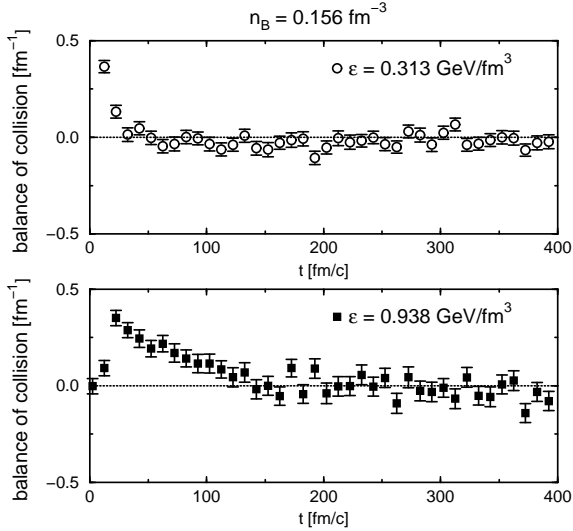}
  \caption{
    The time evolution of the balance of collision frequencies for two
    different energy densities, $\varepsilon=0.313$ GeV/fm$^{3}$ and
    $\varepsilon=0.938$ GeV/fm$^{3}$. In both calculations, the volume
    and baryon number density are $V=64.0$ fm$^{3}$ and
    $n_{B}=0.156$ fm$^{-3}$.  The vertical axes correspond to the
    difference between the frequencies of production processes and
    absorption processes. The fact that the quantities vanish for times
    greater than $t \sim 150$ fm/c indicates that the system 
    satisfies detailed balance for $t \geq 150$ fm/c.
  }
  \label{fig:detail}
\end{figure}

    Figures \ref{fig:pdens} (a) and (b) display the time evolutions
    of particle densities.  These figures show that the system
    approaches the stationary state with time.  The saturation of
    particle densities indicates the realization of chemical
    equilibrium. In order to confirm the detailed balance of each
    reaction processes, we present time evolutions of collision
    frequencies for all kinds of reaction processes in
    Fig. \ref{fig:col}. As seen them in the later stages, the
    collision frequency of the multi-particle production is less than
    $10^{-3}$ (fm/c)$^{-1}$, and the time scale of this process is
    much longer than that of other processes. Violation of reciprocity
    is important only for such long time-scale development.  Actually,
    as shown in Fig. \ref{fig:difnum}, no difference is observed in
    the time evolutions of particle densities when the multi-particle
    production is switched off after $t=150$ fm/c.
    Figure \ref{fig:detail} shows that detailed balance actually holds
    on the time scale of several hundred fm/c.  We conclude that
    chemical equilibrium in our system is realized.
\newpage

    Here we briefly discuss the time scale of the chemical relaxation
    of the system.  Figure \ref{fig:pdens} shows that the number
    density $n(t)$ seems to relax exponentially:
    \begin{eqnarray}
      n(t) & = & (n_{B}-n(\infty))\cdot e^{-\frac{t}{\tau}}+n(\infty)
      \;\;\;\;\;\;\mbox{for}\;\;\;N, \nonumber \\
      n(t) & = &n(\infty)\cdot(1-e^{-\frac{t}{\tau}})
      \;\;\;\;\;\;\mbox{for}\;\;\;\Delta,\;\pi,\;\rho.
      \label{eq:numfit}
    \end{eqnarray}
    Based on this fact, we can easily estimate the relaxation time
    $\tau$ of particle densities that approach chemical equilibrium.
    We give the results for nucleon, pion and $\rho$ densities in
    Table \ref{tab:tau}.  For the nucleon and pion, the obtained
    values of $\tau$ are in the range 7 -- 20 fm/c, and they depend
    on the energy density.  The values of $\tau$ obtained for $\rho$,
    however, are significantly larger.
\begin{table}[h]
  \caption{
    The relaxation time of particle densities for each particle with 
    $n_{B} = 0.156$ fm$^{-3}$.
  }
  \label{tab:tau}
  \begin{tabular}{c|c|c|c} \hline \hline
    $\varepsilon$ [GeV/fm$^{3}$]
    & $N$ (=$\Delta$) [fm/c] & $\pi$ [fm/c] & $\rho$ [fm/c] \\ \hline
    $0.313$ & $ 7.1 \pm 0.4$ & $ 9.2 \pm 0.3$ & $21.2 \pm 0.7$ \\
    \hline
    $0.625$ & $ 9.8 \pm 0.5$ & $13.3 \pm 0.3$ & $44.9 \pm 1.0$ \\
    \hline
    $0.938$ & $10.6 \pm 0.6$ & $18.6 \pm 0.4$ & $68.5 \pm 1.3$ \\
    \hline
  \end{tabular}
\end{table}

    Song and Koch calculated the chemical relaxation time of pions
    in hot hadron gas using the effective chiral
    Lagrangian.\cite{ref:Song1} \ They estimated a chemical relaxation
    time for $\pi$ at $T \sim 150$ MeV as $10$ fm/c, which is close to
    the value obtained the present results.  However,
    Table \ref{tab:tau} shows the existence of several different time
    scales, and some of them are very long.  Though these values may
    depend on the initial conditions of the simulation, our results
    indicate the possibility that hadronic systems have a long
    relaxation time in certain cases.

    \subsection{Thermal equilibration and temperature}
    \label{subsec:temp}
\begin{figure}
  \epsfxsize=10cm
  \epsfbox{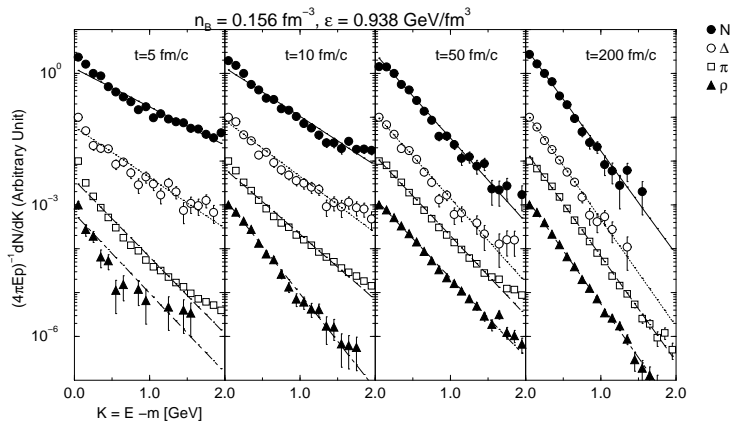}
  \caption{
    Energy distributions of $N_{938}$, $\Delta_{1232}$, $\pi$ and
    $\rho_{770}$ at four different values of time, $t=5$  fm/c,
    $t=10$ fm/c, $t=50$ fm/c and $t=200$ fm/c.  The lines are the
    fitted results that are given by Boltzmann distributions,
    $C\exp(-\beta E)$.  The calculation was done with
    $V=64.0$ fm$^{3}$, $n_{B}=0.156$ fm$^{-3}$ and
    $\varepsilon=0.938$ GeV/fm$^{3}$.
  }
  \label{fig:mom}
\end{figure}
    Figure \ref{fig:mom} displays energy distributions of $N_{938},
    \Delta_{1232},\pi$ and $\rho_{770}$ at $t=5,\;10,\;50$ and $200$ 
    fm/c.  We plot the results as functions of the kinetic energy
    $K = E - m$, so that the horizontal axes for all particle species
    coincide.  The energy distributions approach the Boltzmann
    distribution,
    \begin{equation}
      \frac{dN}{d^{3}\vect{p}} = \frac{dN}{4\pi Ep dE}
      = C\exp(-\beta E),
      \label{eq:Boltz}
    \end{equation}
    as time increases, where $\beta$ is the slope parameter of the
    distribution.  Moreover, the slopes of the energy distributions
    converge to a common value.  These results indicate that
    realization of thermal equilibrium. 

\begin{figure}
  \epsfxsize=9cm
  \epsfbox{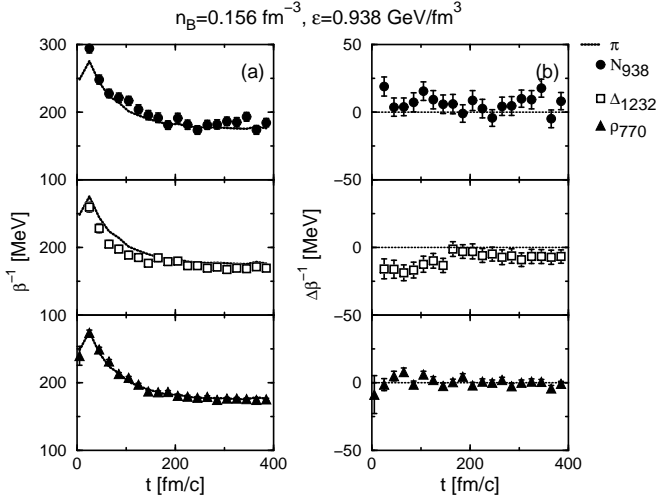}
  \caption{
    (a) The time evolution of the inverse slopes $\beta^{-1}$ for
    $N_{938}$, $\Delta_{1232}$, $\pi$ and $\rho_{770}$ with
    $V=64.0$ fm$^{3}$, $n_{B}=0.156$ fm$^{-3}$,
    $\varepsilon=0.938$ GeV/fm$^{3}$.  The value of $\beta^{-1}$ was
    calculated from the fitting of energy distributions. Here the
    dotted curves represent the time evolution of $\beta^{-1}$ for
    $\pi$. (b) The time evolution of $\Delta\beta^{-1}$ for
    $N_{938}$, $\Delta_{1232}$ and $\rho_{770}$ with
    $V=64.0$ fm$^{3}$, $n_{B}=0.156$ fm$^{-3}$,
    $\varepsilon=0.938$ GeV/fm$^{3}$.  Here $\Delta\beta^{-1}$
    is defined in Eq. (\ref{eq:dbeta}).  The realization of thermal
    equilibrium can be concluded from these graphs.
  }
  \label{fig:slope}
\end{figure}

    Figure \ref{fig:slope} (a) displays the time evolution of the
    quantities $\beta^{-1}$ that were calculated by fitting the energy
    distributions to a Boltzmann distribution.  There, the dotted
    curves correspond to the time evolution of $\beta_{\pi}^{-1}$,
    i.e., the inverse slope of $\pi$.  To confirm the establishment
    of the thermal equilibrium, the difference between the inverse
    slope and that of the pion at time $t$,
    \begin{equation}
      \Delta\beta_{j}^{-1}(t) =
      \beta_{j}^{-1}(t)-\beta_{\pi}^{-1}(t),
      \label{eq:dbeta}
    \end{equation}
    was calculated, where $j$ corresponds to $N_{938}$, $\Delta_{1232}$
    and $\rho_{770}$.  Figure \ref{fig:slope} (b) shows the time
    evolution of these $\Delta\beta_{j}^{-1}$.  From this figure, it
    is seen that the $\Delta\beta_{j}^{-1}$ become zero to within the
    accuracy of the statistics for times later than $150$ fm/c.
    Therefore, we conclude that thermal equilibrium is established
    at about $t = 150$ fm/c; the values of the inverse slope
    parameters of the energy distribution for all particles become
    equal for later times.  Thus we can regard this value as the
    temperature of the system.  For later use, we define the
    temperature of the system as follows:
    \begin{equation}
      T(n_{B},\varepsilon)=\frac{\sum_{j} \beta_{j}^{-1}/
      \sigma_{j}^{2}}{\sum_{j}1/\sigma_{j}^{2}},\;\;\;\;\;\;
      j = N_{938}, \Delta_{1232},\pi\;\; \mbox{and}\;\; \rho_{770},
      \label{eq:temp}
    \end{equation}
    where the $\beta_{j}^{-1}$ are calculated from energy distributions
    that are averaged over time after $t=200$ fm/c, and the
    $\sigma_{j}$ denote the errors of the $\beta_{j}^{-1}$.

  \subsection{Thermodynamic quantities}
  \label{subsec:thermo}
\begin{wrapfigure}{l}{7cm}
  \epsfxsize=7cm
  \epsfbox{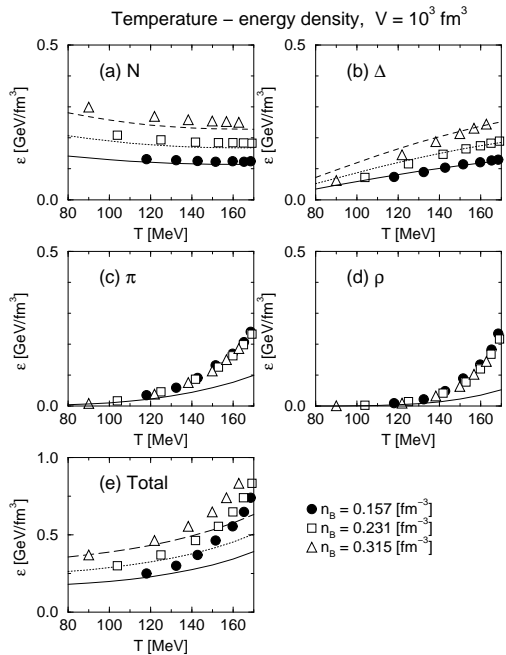}
  \caption{
    The equation of state of a mixed hadron gas at finite temperature
    ($80$ MeV $< T < 170$ MeV) and finite baryon density
    ($0.157$ fm$^{-3} < n_{B} < 0.315$ fm$^{-3}$).  The energy
    densities of (a) $N$, (b) $\Delta$ (the sum of all resonances),
    (c) $\pi$, (d) $\rho_{770}$ and (e) total are plotted as functions
    of the temperature. The curves correspond to the free gas model 
    represented by Eq. (\ref{eq:eoset}).
  }
  \label{fig:et}
\end{wrapfigure}

\begin{figure}[htb]
  \parbox{\halftext}{
  \epsfxsize=7cm
  \epsfbox{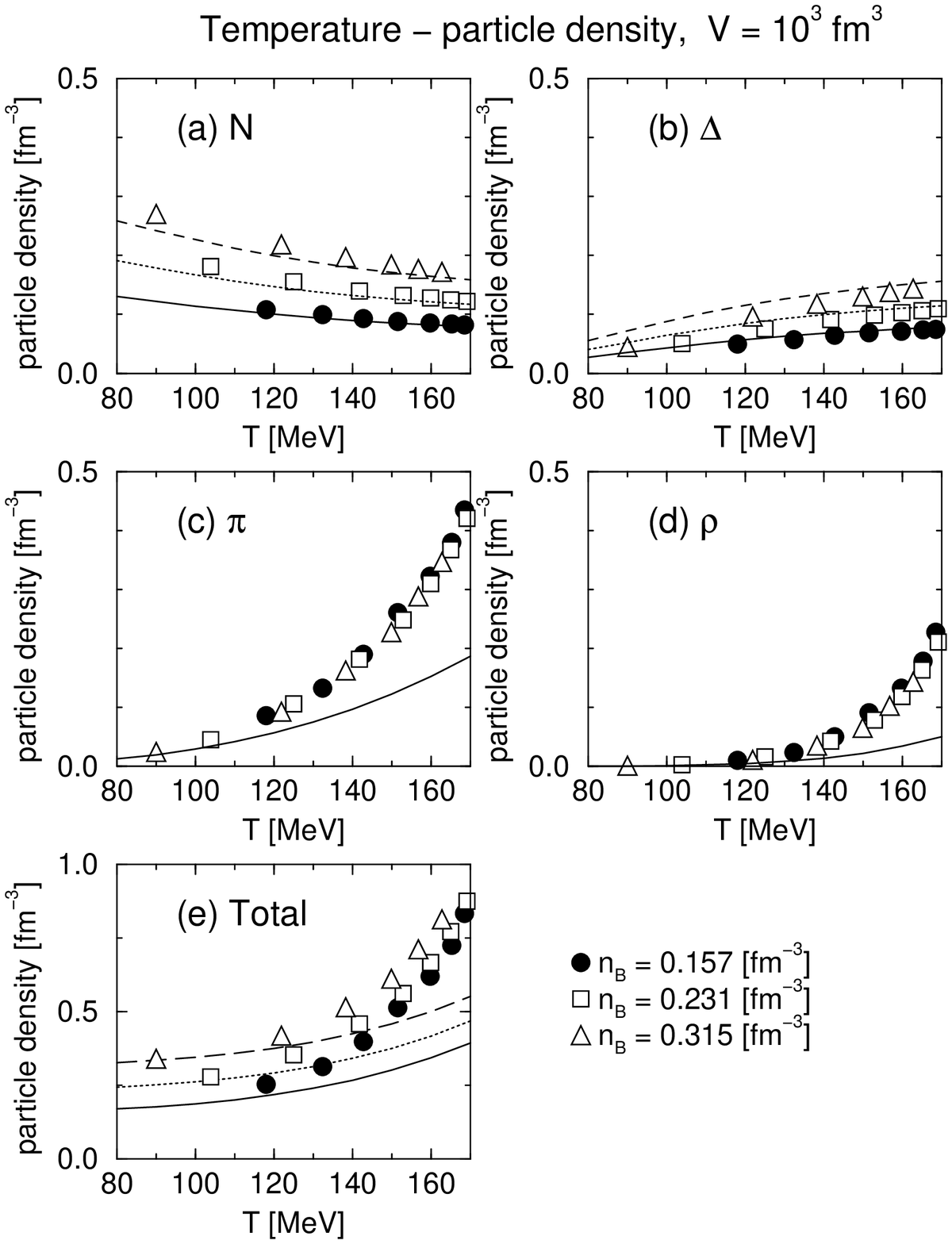}
  \caption{
    Particle densities of (a) $N$, (b) $\Delta$ (the sum of all
    resonances), (c) $\pi$, (d) $\rho_{770}$ and (e) total are plotted
    as functions of the temperature.  The curves correspond to the
    free gas model represented by Eq. (\ref{eq:eospt}).
  }
  \label{fig:tp}
}
  \parbox{\halftext}{
  \epsfxsize=7cm
  \epsfbox{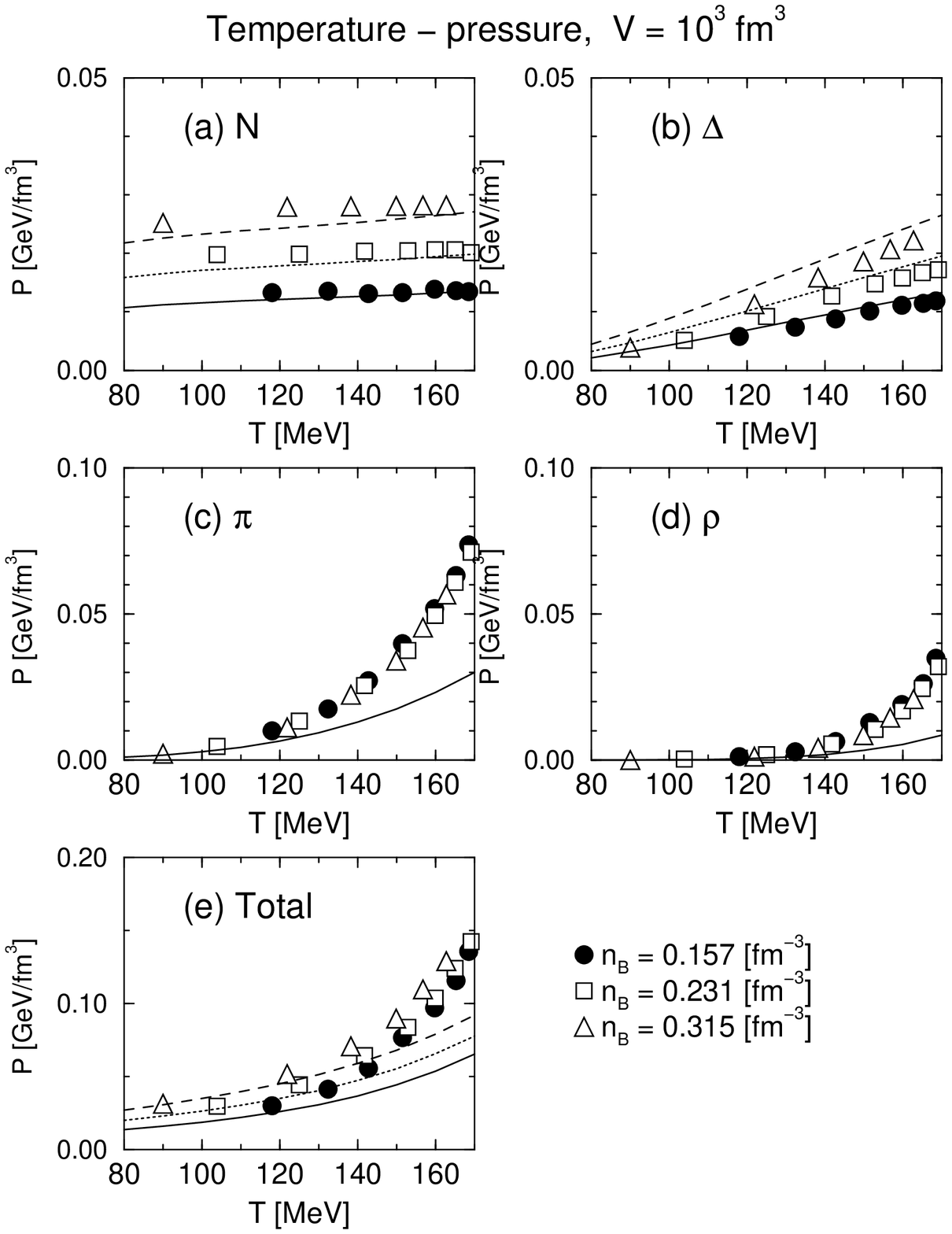}
  \caption{
    Pressures of (a) $N$, (b) $\Delta$ (the sum of all
    resonances), (c) $\pi$, (d) $\rho_{770}$ and (e) total are plotted
    as functions of the temperature.  The curves correspond to the
    free gas model represented by Eq. (\ref{eq:eosprt}).
  }
  \label{fig:tpr}
}
\end{figure}

    Figures \ref{fig:et}--\ref{fig:tpr} show the relations between 
    the temperature and thermodynamic quantities such as energy
    density, $\varepsilon = \frac{1}{V}\sum_{i=1}^{\mbox{\scriptsize all
    particles}}E_{i}$, particle density, and pressure,
      $P = \frac{1}{3V}\sum_{i=1}^{\mbox{\scriptsize all particles}}
      \frac{\vect{p}_{i}\cdot\vect{p}_{i}}{E_{i}}$.  In these figures,
    all curves correspond to the relativistic Fermi-Dirac or
    Bose-Einstein gas with a finite mass width; i.e.,
    \begin{subeqnarray}
    \label{eq:eos}
      \slabel{eq:eoset}
      \varepsilon(T,\mu) & = &
      \sum_{h} d_{h}\int\int\frac{dmd^{3}p}{(2\pi)^{3}}
      \frac{\rho_{R}(m) E}{e^{\frac{E-\mu}{T}} \pm 1},\\
      \slabel{eq:eospt}
      n(T,\mu) & = &
      \sum_{h} d_{h}\int\int\frac{dm d^{3}p}{(2\pi)^{3}}
      \frac{\rho_{R}(m)}{e^{\frac{E-\mu}{T}} \pm 1},\\
      \slabel{eq:eosprt}
      P(T,\mu) & = &
      \sum_{h} d_{h}\int\int\frac{dm d^{3}p}{(2\pi)^{3}}
      \frac{p^{2}}{3E}
      \frac{\rho_{R}(m)}{e^{\frac{E-\mu}{T}} \pm 1},
    \end{subeqnarray}
    where $d_{h}$ is a degeneracy factor and $\rho_{R}(m)$ is the
    mass spectral function, which is given by the Breit-Wigner
    distribution for the resonances. In these calculation, the baryon
    chemical potential $\mu_{B}$ is adjusted to reproduce the total
    baryon number density, whereas the meson chemical potential
    $\mu_{m}$ is fixed to zero.

    In Eq. (\ref{eq:eos}), anti-baryons are ignored, because our event
    generator does not contain anti-baryons.  However, their
    contributions to thermodynamic quantities are negligible
    if the temperature is below 170 MeV, since the ratio of the
    anti-baryon number to the baryon number is suppressed by
    $e^{-\frac{2\mu_{B}}{T}}$.  Even at the smallest value of the
    chemical potential, $\mu_{B} = 250$ MeV ($n_{B}=0.157$ fm$^{-3}$,
    $T=170$ MeV), the factor is about 5\% at most.  Therefore,
    quantitative error caused by ignoring anti-baryons should be only
    about several percent.

    Figures \ref{fig:et}--\ref{fig:tpr} (a) and (b) display the
    equations of state of baryons. In these figures, it is difficult
    to see the difference between our results and those for the
    calculation of the free gas model.  This is the result for the when
    it is under the strict constraint of baryon number conservation,
    which fixes the total number densities of the $N$ and $\Delta$
    particles.  Thus a close look at the fractions of baryon resonances
    is necessary in order to recognize the difference between our
    model and free gas model.

\begin{wrapfigure}{r}{8cm}
  \epsfxsize=8cm
  \epsfbox{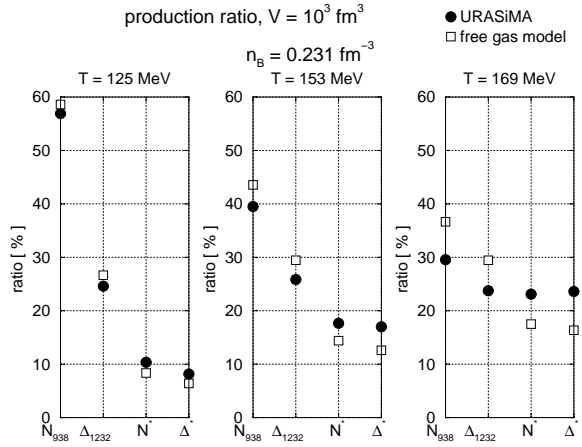}
  \caption{
    The production ratio of $N_{938}$, $\Delta_{1232}$ and other heavy
    resonances, $N^{*}$ and $\Delta^{*}$.  Calculations were done with
    $n_{B}=0.231$ fm$^{-3}$ and temperatures $T=125$ MeV, $153$ 
    MeV, $169$ MeV, respectively.
  }
  \label{fig:prto}
\end{wrapfigure}

    Figure \ref{fig:prto} displays the particle ratios of $N_{938}$,
    $\Delta_{1232}$ and other heavy resonances, $N^{*}$ and
    $\Delta^{*}$.  At lower temperature ($T = 125$ MeV), the
    difference between our results and those of the free gas
    calculations is small. However, at higher temperature ($T > 150$
    MeV), the ratios of light baryons ($N_{938}$ and $\Delta_{1232}$)
    become smaller, and those of heavy resonances, conversely, become
    larger.  Thus we find that the influence of interactions clearly
    appears above $T \sim m_{\pi}$.  Such an enhancement of heavy
    baryons grows as the temperature increases.  The maximum value of
    the enhancement reaches 12--15\% near $T = 170$ MeV. 

    Figures \ref{fig:et}--\ref{fig:tpr} (c) and (d) display the
    equations of state of mesons.  As in the case of baryons, the
    deviation from the free gas model appears at temperatures above
    $T \sim m_{\pi}$.  Above this temperature, meson production is
    large, and the increase of the temperature becomes moderate.

    In a previous study,\cite{ref:Sasa1} \ the saturation of the
    temperature appeared, and there we called it the ``Hagedorn-like
    limiting temperature.''  However, it was an artificial saturation
    of the temperature, because of the luck of the reversal process of
    multi-particle production.  In the calculation of the improved
    URASiMA, this limiting temperature does not appear, and we believe
    that this is an important result of taking detailed balance into
    account.

    Moreover, our results indicate that the equations of state of
    baryons and mesons are closely related through meson-baryon
    interactions, such as $\pi N \rightarrow R$ and their inverse
    processes.  The enhancement of heavy baryon resonances causes the
    increase in the abundances of mesons, and vice versa. Heavy
    resonances readily produce 2$\pi$, and thus the enhancement of
    heavy baryon resonances promotes meson production.  Therefore,
    interactions between mesons and baryons are very important in the
    study of the properties of a mixed hadron gas.

  \subsection{Entropy}
  \label{subsec:Entropy}
\begin{wrapfigure}{l}{6cm}
  \epsfxsize=6cm
  \epsfbox{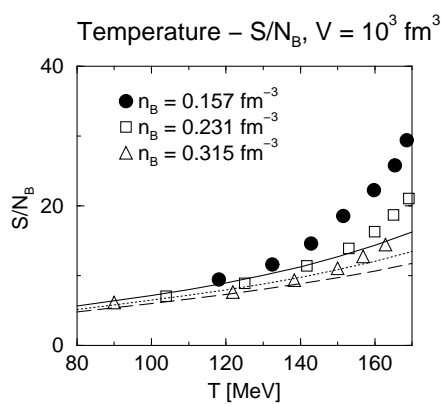}
  \caption{
    The entropy per baryon plotted as a function of the temperature.
    The curves correspond to the free gas model.
  }
  \label{fig:st}
\end{wrapfigure}

    Figure \ref{fig:st} plots the entropy per baryon versus the
    temperature.  Here the curves correspond to free gas calculations.
    In order to define the entropy, we divide the phase space into
    small cells whose volumes are equal to
    $(\hbar c)^{3}$fm$^{3}$(GeV/c)$^{3}$.  We distinguish each cell by
    the index $l$.  The density
    operator of a cell
    $\rho_{l} = \rho(\vect{x}^{l},\vect{p}^{l})$ is defined
    as follows:
    \begin{equation}
      \rho_{l} = \left\{
      \begin{array}{ll}
        1: & \mbox{Particle exists in the {\it l}-th cell.}\\
        0: & \mbox{The {\it l}-th cell is empty.}
      \end{array}
      \right.
    \end{equation}
    The definition of the entropy is given by
    \begin{eqnarray}
      S & = & -\mbox{Tr}\left\{
      \langle\rho_{l}\rangle\ln\langle\rho_{l}\rangle\right\},
      \nonumber \\
        & = & -\sum_{l}^{\mbox{\scriptsize all cells in
        phase space}}\langle\rho_{l}\rangle\ln\langle\rho_{l}\rangle,
      \label{eq:entrpy}
    \end{eqnarray}
    where $\langle\rho_{l}\rangle$ is the ensemble average of the
    density operator of a cell,
    \begin{equation}
     \langle\rho_{l}\rangle 
     =\frac{1}{\mbox{number of ensemble states}}
     \sum_{\mbox{{\scriptsize ensemble}}}\rho_{l},
    \end{equation}
    and the trace constitutes the average over phase space.
    Figure \ref{fig:st} shows that the hadron gas has a larger value of
    $S/N_{B}$ in the present model than in the free gas model for
    $T > m_{\pi}$, because a large part of the energy of the system
    goes into particle (entropy) production.  Moreover, our results
    show a clear dependence on the baryon density.  Thus, the
    frequently assumed ansatz that $S/N_{B}$ is insensitive to $n_{B}$
    seems unreasonable.  Therefore, our results indicate that the
    free gas model provides a poor description for these quantities
    above $T = m_{\pi}$.  Hence, we should be more careful in using a
    free gas model in the interpretation of ultra-relativistic heavy
    ion experiments, even if some corrections, such as the excluded
    volume effect, are considered.\cite{ref:Mun1}\tocite{ref:Mun3}

    Furthermore, some models seem insufficient in counting the
    dynamical degrees of freedom per baryon.  For example, EOS based on
    the $\sigma$-$\omega$ model\cite{ref:Ris1} predicts
    $S/N_{B} \sim 5$ (see Ref.~\citen{ref:Rei1}) independently of the
    temperature, while our result gives $S/N_{B} \sim 18.6 \pm 0.2$
    at $T = 150$ MeV and $29.2 \pm 0.1$ at $T = 170$ MeV for the
    normal nuclear matter density.

  \subsection{Energy density dependence of the pressure}
  \label{subsec:ep}
\begin{wrapfigure}{r}{8cm}
  \epsfxsize=8cm
  \epsfbox{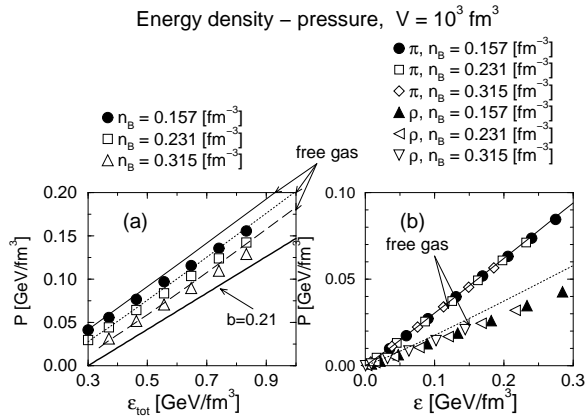}
  \caption{
    (a) Pressures are plotted as functions of the energy density.
    Free gas results and a line with slope 0.21 are also shown.
    (b) Partial pressures of mesons as functions of their partial
    energy densities.
  }
  \label{fig:ep}
\end{wrapfigure}

    From Fig. \ref{fig:ep} (a), we find that our results and those of
    the free gas calculations exhibit almost a linear dependence of
    the pressure on the energy density within the range of the present
    study.  Our results can be fitted by the following function with
    parameters $b$ and $\varepsilon_{0}$:
    \begin{equation}
      P(\varepsilon) = b \cdot (\varepsilon - \varepsilon_{0}).
      \label{eq:ep}
    \end{equation}
    The values of these parameters at three different baryon number
    densities are shown in Table \ref{tab:epfit}.
  
    To clarify the situation further, we now investigate partial
    pressures.  As shown in Figs.~\ref{fig:et}--\ref{fig:tpr},
    pressures and energy densities of baryons as functions of the
    temperature do not deviate from their forms for the free gas.
    Contrastingly, the partial pressure of the mesons shows a clear
    deviation from that of the free gas model.
    In Fig. \ref{fig:ep} (b) the partial pressures of mesons are plotted
    as functions of the corresponding partial energy densities.  It is
    interesting that pions behave like a free gas with regard to
    this quantity.  On the other hand, the slope of the $\rho$ meson
    is smaller in the present case than in the free gas.  This result
    indicates that the $\rho$ meson and its interactions play an
    important role in determining the thermodynamic properties of a
    hadron gas.
\begin{wraptable}{r}{\halftext}
  \caption{
    Fitting parameters for Fig. \ref{fig:ep}.
  }
  \label{tab:epfit}
  \begin{tabular}{c|c|c} \hline \hline
    $n_{B}$ [fm$^{-3}$]  & $b$ & $\varepsilon_{0}$
    [GeV/fm$^{3}$] \\
    \hline
    $0.157$ & $0.2171 \pm 0.0009$ & $0.1163 \pm 0.0026$ \\ \hline
    $0.231$ & $0.2122 \pm 0.0008$ & $0.1633 \pm 0.0023$ \\ \hline
    $0.315$ & $0.2083 \pm 0.0007$ & $0.2188 \pm 0.0021$ \\ \hline
  \end{tabular}
\end{wraptable}

  \subsection{Finite size effect}
  \label{subsec:vol}
\begin{figure}
  \epsfxsize=9cm
  \epsfbox{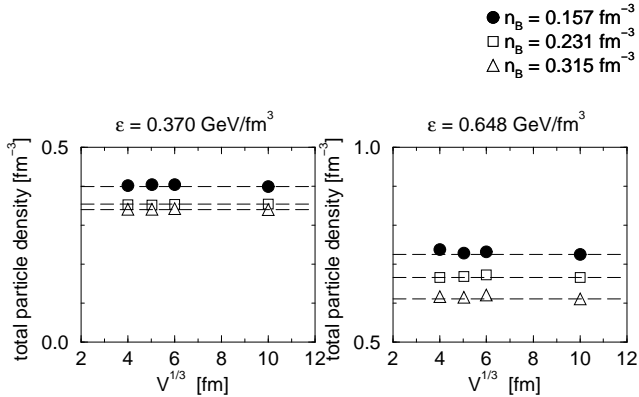}
  \caption{
    Volume dependence of the total particle density for $\varepsilon
    = 0.370$, $0.648$ GeV/fm$^{3}$.
  }
  \label{fig:Vpn}
\end{figure}

    Finally, we check the finite size effect of our calculation.  To
    see this, we prepared four different sizes of the boxes,
    $64.0$ fm$^{3}$, $128.0$ fm$^{3}$, $216.0$ fm$^{3}$ and
    $1000.0$ fm$^{3}$. The volume dependence of
    the total particle density is shown in Fig. \ref{fig:Vpn}. We were
    not able to find any differences among the four sizes of the box.
    Other thermodynamic quantities yield the same result.  As concerns
    thermodynamic quantities, we can regard $64.0$ fm$^{3}$ as a
    sufficiently large spatial size.

\begin{figure}
  \epsfxsize=10cm
  \epsfbox{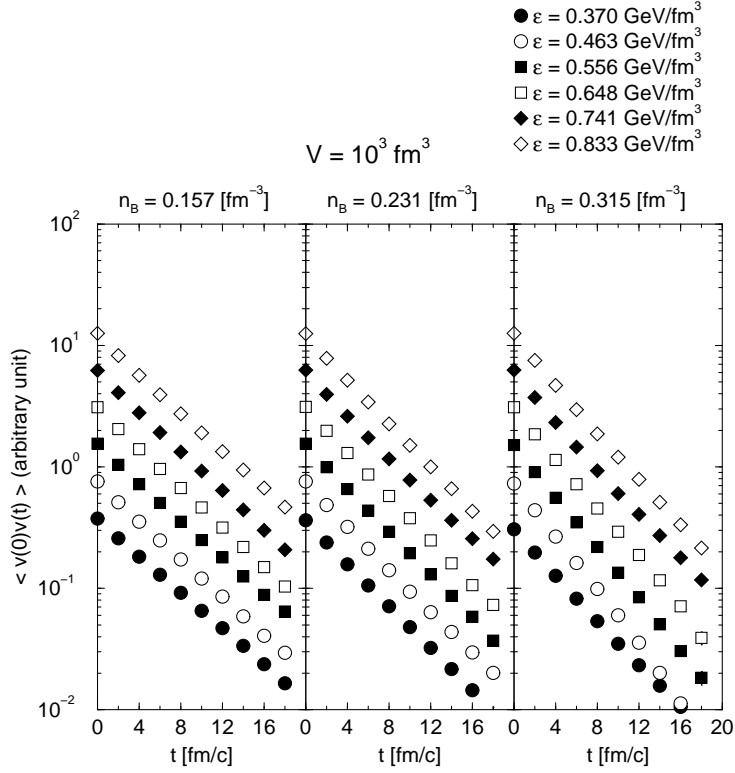}
  \caption{
    Velocity correlation of the baryons as a function of time.
    The normalization of the data is arbitrary.
  }
  \label{fig:vvcor}
\end{figure}

\begin{figure}
  \epsfxsize=10cm
  \epsfbox{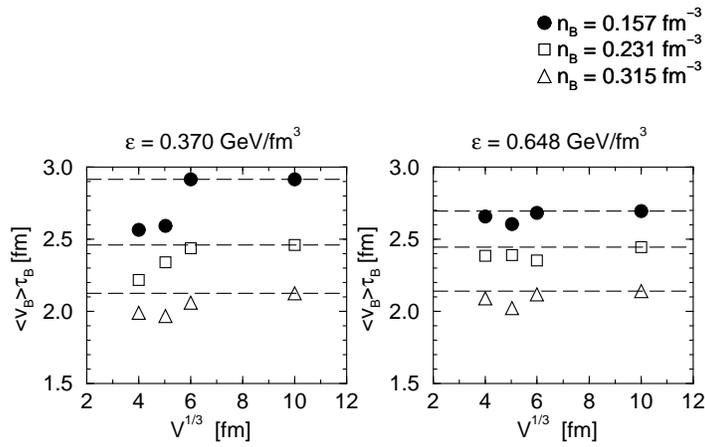}
  \caption{
    The relaxation time multiplied by average velocity as a function
    of the volume of the system.
  }
  \label{fig:Vcorlenb}
\end{figure}

    We also compared relaxation lengths for various spatial box sizes.
    In a previous paper, we investigated the baryon diffusion
    constant.\cite{ref:Sasa2} \ First, the fluctuation-dissipation
    theorem tells us that diffusion constant $D$ is given by the
    current (velocity) correlation,\cite{ref:Kubo1}
    \begin{equation}
      D =\frac{1}{3}
      \int_{0}^{\infty}\langle\vect{v}(t)\cdot \vect{v}(t+t')
      \rangle dt'.
      \label{eqn;fdt}
    \end{equation}
    In our definition, the correlation function $\langle\cdots\rangle$
    is given by
    \begin{equation}
      \langle\cdots\rangle = \frac{1}{\mbox{number of ensemble
       states}}
      \sum_{\mbox{{\scriptsize ensemble}}}
      \frac{1}{\mbox{number of baryons}}
      \sum_{\mbox{{\scriptsize baryons}}} \cdots .
      \label{eqn;avg}
    \end{equation}
    The correlation functions are damped exponentially with time
    (see Fig. \ref{fig:vvcor}):
    \begin{equation}
      \langle\vect{v}(t)\cdot \vect{v}(t+t')\rangle \propto
      \exp{\left (- \frac{t'}{\tau_{B}}\right )}.
      \label{eqn;rlx}
    \end{equation}
    Thus the diffusion constant can be rewritten in the simple form
    \begin{equation}
      D = \frac{1}{3}
      \langle\vect{v}(t)\cdot \vect{v}(t)\rangle \tau_{B},
      \label{eqn;difcon}
    \end{equation}
    where $\tau_{B}$ is the relaxation time of the baryon current.
    Figure \ref{fig:Vcorlenb} shows the volume dependence of
    $\tau_{B}\cdot \langle v_{B} \rangle$, where
    $\langle v_{B} \rangle$ denotes the average speed of baryons,
    and thus $\tau_{B}\cdot \langle v_{B} \rangle$ has dimensions
    of length.  Though this quantity has a clear dependence on the
    size of the box at low energy density ($\varepsilon = 0.370$
    GeV/fm$^{3}$), this box size dependence seems to disappear for
    volumes larger than about $6^{3}$ fm$^{3}$.  In this work, to be
    safe, we used a box of volume $V = 1000$ fm$^{3}$.  Therefore, the
    result in this paper can be considered free of any box-size effect.

  \section{Summary}
  \label{sec:summary}
    In this paper, we studied the equilibration and the equation of
    state of hot ($80$ MeV $< T < 170$ MeV) hadron gas with finite
    baryon number density ($0.157$ fm$^{-3}$ $< n_{B}<0.315$ fm$^{-3}$)
    by using the event generator URASiMA, which maintains detailed
    balance in the practical sense in the hadronic energy scale.  We
    performed molecular dynamical calculations for a system of hadrons
    in a box with periodic boundary conditions.  The energy density
    and baryon number in our simulation correspond to those of a hot
    hadron fluid, which is believed to be produced in high energy
    nuclear collisions.  Our results for thermodynamic quantities can
    be related to experimental data through statistical models and
    hydrodynamic models.\cite{ref:Nona1}

    Collision frequencies and particle densities exhibit saturation as
    the time evolution proceeds.  These results indicate that the
    system reaches chemical equilibration.  We also studied the
    thermal equilibration of the system from the time evolution of
    the inverse slopes of the energy distributions. To confirm that
    thermal equilibrium is established, it was demonstrated that the
    slope parameters of $N_{938}$, $\Delta_{1232}$, $\pi$ and
    $\rho_{770}$ become almost identical for times greater than
    $t \sim 150$ fm/c.  Thus the temperature of the system can be
    defined after this time.

    After the equilibration, thermodynamic quantities were evaluated,
    and the equation of state was investigated. Energy densities,
    number densities, pressures and entropies per baryon were plotted
    as functions of the temperature.  Deviations from the free gas
    model were manifestly observed above $T \sim m_{\pi}$.  Above this
    temperature, the excitation of heavy baryon resonances were found
    to be enhanced, and meson production was found to be significant.
    Those effects suppress the increase of the temperature, but the
    saturation of the temperature, as in the case of Hagedorn's
    limiting temperature, never occurs.  These notable differences
    from previous works\cite{ref:Bel1,ref:Sasa1} are important
    results of the proper maintenance detailed balance between
    production processes and absorption processes.  For a reliable
    simulation of a statistical system, detailed balance is
    essentially important.  In our study, we found large values of 
    the entropy per baryon. This depends strongly on the baryon
    density above $T \sim m_{\pi}$.  The pressures exhibit linear
    dependences on the energy densities within the range of present
    study.  Such behavior was analyzed in detail by looking at
    relations between partial pressures and partial energy densities
    of mesons.  We find that the $\rho$ meson and its interactions
    play an important role in the thermodynamic properties of the
    hadron gas.  

    Because the temperature in this investigation (80--170 MeV) is
    much lower than the mass of a $K\bar{K}$ pair, the hadron gas is
    mainly composed of non-strange particles.  For this reason,
    strange particles are not considered in our model.  However,
    in the early stages of the AA collision, the energy density can
    be very high, and the strangeness degree of freedom should play
    an important role.  Including strange particles in our model is
    the next task, and we are now in process of doing so.

    The investigation reported in this paper was focused on
    ``hadronic scale'' energy densities and baryon number densities.
    However, it would be interesting to perform calculations at
    higher energy densities and baryon number densities.  Our EOS
    without a phase transition can work as a helpful reference to the
    equation of state with a QGP phase transition.

  \section*{Acknowledgements}
     This work was carried out under the direction of Professor
     Osamu Miyamura, Hiroshima University, who passed away on July
     10th, 2001.  We would like to thank Professor Osamu Miyamura
     for his dedicated supervision; without his continuous
     encouragement this work would never have been finished.  Dr. Shin
     Muroya and Dr. Chiho Nonaka carefully read the manuscript and
     made a number of helpful comments.  Professor Atsushi Nakamura
     and Dr. Schin Dat\'e gave us constructive advice throughout the
     writing of the paper.  Calculations were done at the Institute
     for Nonlinear Science and Applied Mathematics (INSAM),
     Hiroshima University.

\appendix
\section{Event Generator URASiMA}
\subsection{The search for would-be collision points}
    \label{sec:colp}
    After a collision at the space-time point $x_{0\,i}^{\mu} =
    (t_{0\,i},\vect{x}_{0\,i})$, the trajectory of the $i$-th particle
    is given by the straight line
    \begin{equation}
      x_{i}^{\mu} = (t_{i},\vect{x}_{i}) =
      \frac{p_{i}^{\mu}}{m_{i}}\xi_{i} + x_{0\,i}^{\mu},
      \label{eq:xmu}
    \end{equation}
    where $p_{i}^{\mu} = (E_{i},\vect{p}_{i})$ and $m_{i}$
    are the momentum and the mass of the particle after the collision,
    and $\xi_{i}$ represents the proper time:
    \begin{equation}
      \xi_{i} \equiv \frac{m_{i}}{E_{i}}(t_{i}-t_{0\,i}).
      \label{eq:xi}
    \end{equation}
\begin{wrapfigure}{r}{5cm}
  \epsfxsize=5cm
  \epsfbox{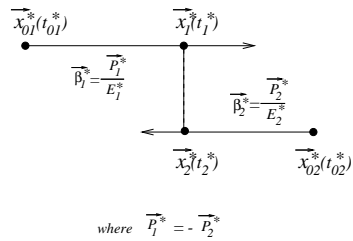}
  \caption{
    The condition for which the collision occurs in the CM frame of
    two particles. In this frame, the momentum vector
    $p^{*}_{1}$ $\|$ $p^{*}_{2}$ and the difference
    between positions
    $x^{*}_{1}(t^{*}_{1})-x^{*}_{2}(t^{*}_{2})$
    become perpendicular at the point of nearest approach.  In this
    situation, a collision occurs if Eq. (\ref{eq:b}) is satisfied.
  }
  \label{fig:orbit}
\end{wrapfigure}
    Equation (\ref{eq:xmu}) holds until the next interaction occurs.

    We adopt a hard sphere approximation for binary collisions;
    that is, a collision occurs whenever the condition
    \begin{equation}
      \pi \vect{b}^{*\,2} < \sigma
      \label{eq:b}
    \end{equation}
    is satisfied, where $|\vect{b}^{*}|$ is the impact parameter in
    the CM frame of two particles, and $\sigma$ is the total cross
    section. For the manifestly relativistic formulation, it is
    important to express the collision condition (\ref{eq:b}) in a
    covariant way.  For this purpose, we consider the CM frame of two
    particles, 1 and 2 (see Fig. \ref{fig:orbit}).  When two particles
    are at the points of their closest approach, the impact parameter
    $\vect{b}^{*} \equiv \vect{x}^{*}_{1}(t^{*}_{1})
    -\vect{x}^{*}_{2}(t^{*}_{2})$ and the time $t^{*}_{c}$ are
    defined as
    \begin{eqnarray}
      (\vect{x}^{*}_{1}(t^{*}_{1})-\vect{x}^{*}_{2}(t^{*}_{2}))
      \cdot \vect{p}^{*}_{j} & = & 0,
      \;\;\;\;\;\; (j=1,2) \nonumber \\
      t^{*}_{1} = t^{*}_{2} \equiv t^{*}_{c},
      \label{eq:ccond1}
    \end{eqnarray}
    where $t^{*}_{c}$ denotes the time at which a collision may occur.
    Here, we define the total momentum $Q^{\mu *}$ and the difference
    of momenta $K^{\mu *}$ as
    \begin{eqnarray}
      Q^{\mu *} & \equiv & p_{1}^{\mu *}+p_{2}^{\mu *}
      = (E_{1}^{*}+E_{2}^{*},0), \nonumber \\
      K^{\mu *} & \equiv & p_{1}^{\mu *}-p_{2}^{\mu *}
      = (E_{1}^{*}-E_{2}^{*},\vect{p}^{*}_{1}-\vect{p}^{*}_{2}).
      \label{eq:QK}
    \end{eqnarray}
    Using $Q^{\mu *}$ and $K^{\mu *}$, Eq. (\ref{eq:ccond1}) can be
    rewritten as
    \begin{eqnarray}
      (\vect{x}^{*}_{1}(t^{*}_{1})-\vect{x}^{*}_{2}(t^{*}_{2}))
      \cdot \vect{K}^{*} & = & 0, \nonumber \\
      (t^{*}_{1} - t^{*}_{2})Q^{0 *} & = & 0.
      \label{eq:ccond2}
    \end{eqnarray}
    The invariant expressions of Eq. (\ref{eq:ccond2}) are easily
    obtained as follows:
    \begin{eqnarray}
      (x_{1}-x_{2}) \cdot Q & = & 0,\nonumber \\
      (x_{1}-x_{2}) \cdot K & = & 0.
      \label{eq:ccond3}
    \end{eqnarray}
    Replacing $x_{1}$ and $x_{2}$ in Eq. (\ref{eq:ccond3}) by
    their forms in Eq. (\ref{eq:xmu}), we find the equations
    \begin{eqnarray}
      \frac{p_{1} \cdot Q}{m_{1}}\xi_{1}-
      \frac{p_{2} \cdot Q}{m_{2}}\xi_{2}+
      X_{0} \cdot Q & = & 0, \nonumber \\
      \frac{p_{1} \cdot K}{m_{1}}\xi_{1}-
      \frac{p_{2} \cdot K}{m_{2}}\xi_{2}+
      X_{0} \cdot K & = & 0, 
      \label{eq:ccond4}
    \end{eqnarray}
    where $X^{\mu}_{0} = x^{\mu}_{0\,1}-x^{\mu}_{0\,2}$.  The
    solutions of these equations are easily obtained as
    \begin{equation}
      \left( 
      \begin{array}{c}
        \xi_{1} \\
        \xi_{2}
      \end{array}
      \right) = \frac{-1}{J}\left(
      \begin{array}{cc}
       -\frac{p_{2} \cdot K}{m_{2}} & 
       \frac{p_{2} \cdot Q}{m_{2}} \\
       -\frac{p_{1} \cdot K}{m_{1}} & 
       \frac{p_{1} \cdot Q}{m_{1}}
      \end{array}
      \right)\left(
      \begin{array}{c}
        X_{0} \cdot Q \\
        X_{0} \cdot K \\
      \end{array}
      \right),
      \label{eq:xi1xi2}
    \end{equation}
    where $J$ is defined as
    \begin{eqnarray}
      J & \equiv &
      -\frac{(p_{1}\cdot Q)(p_{2}\cdot K)-(p_{1}\cdot K)(p_{2}\cdot Q)}
      {m_{1}m_{2}},
      \nonumber \\
        & = &
       -2\frac{(p_{1}\cdot p_{2})^{2}-(m_{1}m_{2})^{2}}{m_{1}m_{2}}.
      \label{eq:J}
    \end{eqnarray}
    Using Eqs. (\ref{eq:xmu}) and (\ref{eq:xi1xi2}), the invariant
    expression for the impact parameter is given by
    \begin{eqnarray}
      b^{2} & \equiv & (x_{1}-x_{2})^{2} \nonumber \\
      & = & X^{2}_{0} - \frac{(X_{0} \cdot Q)^{2}}{Q^{2}} -
      \frac{ [ (X_{0} \cdot K) -
      \frac{(K \cdot Q)(X_{0} \cdot Q)}{Q^{2}} ]^{2} }
      {(K - \frac{(K \cdot Q)}{Q^{2}} Q )^{2}}.
      \label{eq:imp}
    \end{eqnarray}
    This expression enables us to specify the collision point in terms
    of momenta and space-time coordinates of the starting points (the
    previous collision points).  In the simulation, all candidates for
    collision are searched for using Eq. (\ref{eq:imp}), and the
    earliest collision in the rest frame of the box is generated.

\subsection{Absorption cross sections}
    \label{sec:abs}
    By using the reciprocity of the S matrix, the absorption cross
    section can be related to the production cross section
    \begin{equation}
      \sigma_{N''R \rightarrow NN'}=
      \frac{2}{g_{R}}\frac{1}{1+\delta_{NN'}}
      \frac{p_{f}^{2}}{p_{i}^{2}}
      \sigma_{NN' \rightarrow N''R},
     \label{eq:detailed} 
    \end{equation}
    with $\delta_{NN'}$ the Kronecker $\delta$, which is equal to 1 
    when the final state nucleons belong to the same state.
    The relation Eq. (\ref{eq:detailed}) is limited to particles with
    definite masses.  Extension of Eq. (\ref{eq:detailed}) to take
    into account the width of resonance mass is straightforward,
    \cite{ref:Dan1}\tocite{ref:Li1} \ and we have
    \begin{equation}
      \sigma_{N''R \rightarrow NN'} = \frac{2}{g_{R}}
      \frac{m_{R}p^{2}_{f}}{1+\delta_{NN'}}
      \frac{ \sigma_{NN' \rightarrow N''R}}
      {p_{i}\int^{\sqrt{s}-m_{N}}_{m_{\pi}+m_{N}}
      \frac{dm^{'}_{R}}{2\pi}m^{'}_{R}\rho_{R}(m^{'}_{R})p^{'}_{i} },
      \label{eq:EDB2}
    \end{equation}
    where $\rho_{R}(m^{'}_{R})$ is the mass distribution function,
    which is given by the Breit-Wigner distribution.
    
\subsection{Multi-particle productions}
    \label{sec:mul}
\begin{wrapfigure}{l}{6cm}
  \epsfxsize=6cm
  \epsfbox{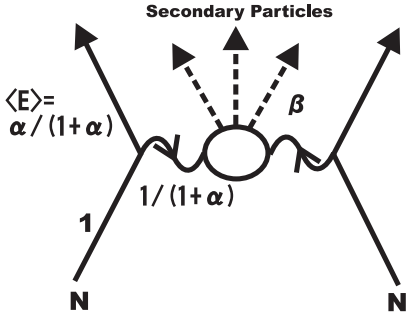}
  \caption{ 
    The diagram of the multi-particle production process for an
    $N + N$ inelastic collision.  Here $\langle E \rangle$ denotes
    the average of the energy fraction that the primary nucleon
    carries after the collision. This quantity is related to the
    value of $\alpha$, which is one of the adjustable parameters.
    The remainder of the energy fraction is consumed by the
    production of secondary particles, whose momenta are determined
    according to the distribution characterized by $\beta$.
  }
  \label{fig:diagram}
\end{wrapfigure}
    In high energy hadron-hadron collisions, {\it n}-particle
    ($n \geq 3)$ productions may occur.  In our model, such 
    multi-particle productions are treated as direct processes.  Such
    processes are very important in the early stages, where energetic
    particles dominate the system.  In the present simulation, initial
    states include many energetic nucleons, and multi-particle
    production takes place frequently.  In the later stages, however,
    they seldom occur, and their effect is negligibly small. 

    In URASiMA, a multi-particle production process is realized by use
    of the multi-chain model,\cite{ref:Cape1}\tocite{ref:Kan1} \ where
    the exchange of a hadronic chain causes the direct emission of
    multipions.  Figure \ref{fig:diagram} displays a diagram of $N + N$
    inelastic collisions.  In this figure, nucleons exchange a chain,
    which produces secondary particles.  Such multi-particle production
    is specified by parameters as follows.  The probability
    distribution of a light-like longitudinal momentum fraction $x_N$
    that is carried by nucleons after collisions is given by
    \begin{equation}
      P(x_{N}) = \alpha \cdot x^{\alpha -1}_{N}.
      \label{eq:lead}
    \end{equation}
\begin{wraptable}{r}{\halftext}
  \caption{
    Parameters of multi-particle production.
  }
  \label{tab:mcmp}
  \begin{tabular}{c|c|c} \hline \hline
     $\;\;\;\;\;\;\alpha\;\;\;\;\;\;$
   & $\;\;\;\;\;\;\beta\;\;\;\;\;\;$
   & $\;\;\;\;\;\;\tau_{0}$ [fm/c]$\;\;\;\;\;\;$ \\ \hline
        1        &   1.0    &   1.0  \\ \hline
  \end{tabular}
\end{wraptable}
    The average energy that is carried by nucleons is
    $\alpha/(\alpha + 1)$ times the initial energy. The remainder,
    which is on average equal to $1/(\alpha + 1)$, is consumed by the
    production of secondary particles.

    In the C.M. frame of produced particles, the longitudinal momentum
    of the secondary particle behaves according to the distribution
    \begin{equation}
      \frac{dN}{dy} \propto
      \left (1 - \frac{2m_{T}\cosh{y}}{\sqrt{\hat{s}}}
      \right )^{\beta},
      \label{eq:sec}
    \end{equation}
    with $m_{T}$ and $y$ the transverse mass and the rapidity of
    the secondary particle.  The quantity $\sqrt{\hat{s}}$ stands for
    the energy deposited in the blob of the diagram.

    Transverse momentum distributions of primary and secondary
    particles are given by gamma distributions:
    \begin{equation}
      \frac{dN_{j}}{dp_{T}} \propto p_{T} e^{-B_{j}\cdot p_{T}}.
      \;\;\;\;\; (j = \mbox{nucleon},\;\;\pi,\;\;\mbox{etc}.)
      \label{eq:ptdist}
    \end{equation}
    Here $p_{T}$ denotes the transverse momentum, and the slope
    parameter $B_{j}$ is another parameter of the model.

    Secondary particles propagate freely without any interaction during
    the formation time $\tau$ after the emission,
    \begin{equation}
      \tau = \gamma \cdot \tau_{0} = \frac{E}{m} \cdot \tau_{0},
      \label{eq:form}
    \end{equation}
    where the proper formation time $\tau_{0}$ is also one of the
    parameters of URASiMA. Values of the parameters $\alpha$, $\beta$
    and $\tau_{0}$ are listed in Table \ref{tab:mcmp}.

\subsection{Comparisons with data of nucleus-nucleus collisions
    at high energies}
    \label{sec:exp}
    In order to examine the descriptive power of the event generator
    URASiMA-B, we made comparisons with the data of high-energy
    nuclear collision experiments.  In this case, the initial state is
    given by the energetic nucleons forming the nuclei in free space.
    All parameters of the model are tuned to reproduce the data of
    hadron-hadron collisions.\cite{ref:pdg1}\tocite{ref:Zab1}

\begin{figure}
  \epsfxsize=15cm
  \epsfbox{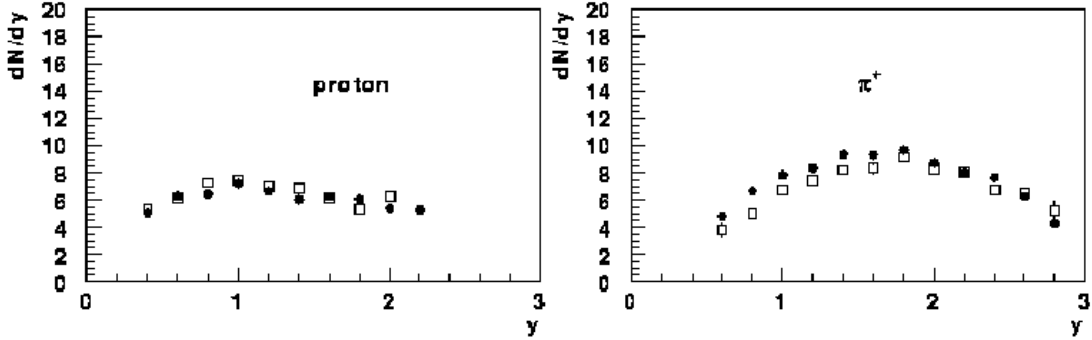}
  \caption{
    A comparison between experimental results and our simulations.
    We plot the rapidity distributions of the proton (left) and
    $\pi$ (right) for Si + Al central collisions at 14.6 GeV/nucleon
    obtained in the E802 collaboration.  Here, the squares denote the
    results of our simulation, and the filled circles denote the
    experimental results.
  } 
  \label{fig:rap}
\end{figure}

\begin{figure}
  \epsfxsize=15cm
  \epsfbox{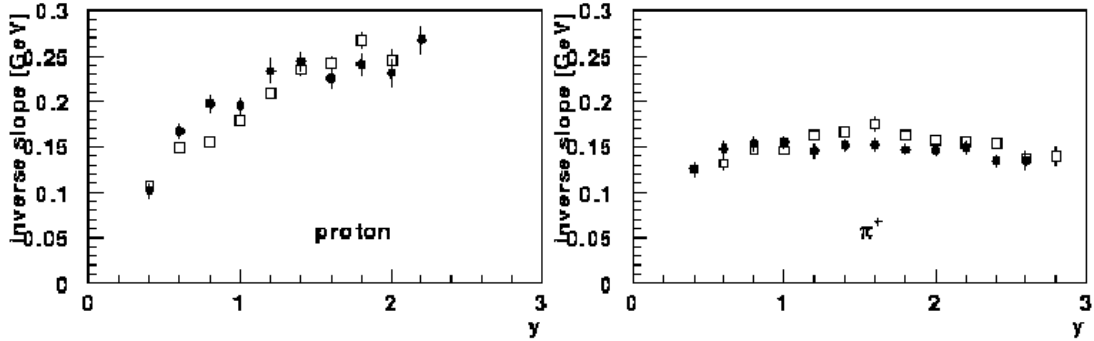}
  \caption{
    A comparison between experimental results and our simulations.
    We plot inverse slopes of transverse momentum distributions of the
    proton (left) and $\pi$ (right) for Si + Al central collisions at
    14.6 GeV/nucleon obtained in the E802 collaboration. Here, the
    squares denote the results of our simulation, and the filled
    circles denote the experimental results.
  }
  \label{fig:ptslope}
\end{figure}

    The results of our simulations are compared with the experimental
    data of the E802 collaboration for the Si + Al central collision
    at $14.6$ GeV/nucleon.\cite{ref:Abbo1} \ Figure \ref{fig:rap}
    displays the rapidity distributions for the proton (left) and
    $\pi^{+}$ (right), where the squares denote the results of our
    simulation and the filled circles denote the experimental results.
    Figure \ref{fig:ptslope} shows the rapidity dependence of the
    inverse slopes of transverse momentum distributions for the proton
    (left) and $\pi^{+}$ (right).  In both figures, URASiMA-B
    reproduces global features of experimental data quite well.
\newpage

  
\end{document}